\documentclass{article}
\usepackage{amsfonts}
\usepackage{amsmath}
\usepackage{amssymb}
\usepackage{amsthm}
\usepackage{amstext}
\usepackage{float}
\usepackage{graphicx}
\input xy
\xyoption{all}
\newcommand{\Rho}{\mathrm{P}}
\newcommand{\sech}{\operatorname{sech}}
\newcommand{\KP}{\scriptscriptstyle \textrm{KP}}
\newcommand{\mKP}{\scriptscriptstyle \textrm{mKP}}
\title{Matrix solutions of a noncommutative KP equation and a noncommutative mKP equation}
\author{ C. R. Gilson, J. J. C. Nimmo and C. M. Sooman \\
Department of Mathematics, \\
University of Glasgow \\
Glasgow G12 8QW, UK}
\date{}

\begin{document}

\maketitle

\begin{abstract}
Matrix solutions of a noncommutative KP and a
noncommutative mKP equation which can be expressed as
quasideterminants are discussed. In particular, we
investigate interaction properties of two-soliton solutions.
\end{abstract}

\section{Introduction}
A considerable amount of literature exists concerning noncommutative integrable systems. This includes noncommutative versions of the Burgers equation, the KdV equation, the KP equation, the mKP equation and the sine-Gordon equation \cite{gilson-2008-41, MR2114103, MR2379500, Dimakis:Muller-Hoissen2006, MR2095463, MR1858198, claire:jon, MR2025480, MR2186388, MR1805720}.  These equations can often be obtained by simply removing the assumption that the dependent variables and their derivatives in the Lax pair commute.

This paper is concerned with a noncommutative KP equation (ncKP) and a noncommutative mKP equation (ncmKP) \cite{MR2095463, MR1752088}.  The Lax pair for ncKP is given by
\begin{align}
 L_{\KP} &= \partial_x^2 + v_x - \partial_y, \\
 M_{\KP} &= 4 \partial_x^3 + 6 v_x \partial_x + 3 v_{xx} + 3v_y + \partial_t.
\end{align}
Both $L_{\KP}$ and $M_{\KP}$ are covariant under the Darboux transformation \\ $G_{\theta} = \theta \partial_x \theta^{-1}$, where $\theta$ is an eigenfunction for $L_{\KP},M_{\KP}$.  From the compatibility condition $[L_{\KP},M_{\KP}]=0$ we obtain a noncommutative version of the KP equation:
\begin{align}\label{ncKP_eqn}
    (v_t + 3v_xv_x + v_{xxx})_x + 3v_{yy} + 3[v_x,v_y] = 0,
\end{align}
where $u=v_x$.  For ncmKP, the Lax pair \cite{MR2095463} is given by
\begin{align}
L_{\mKP} &= \partial_x^2 + 2 w \partial_x - \partial_y, \\
M_{\mKP} &= 4 \partial_x^3 + 12 w \partial_x^2 + 6(w_x  + w^2 + W)\partial_x + \partial_t.
\end{align}
Both $L_{\mKP}$ and $M_{\mKP}$ are covariant under the Darboux transformation \\ $G_{\theta} = ((\theta^{-1})_x)^{-1} \partial_x \theta^{-1}$, where $\theta$ is an eigenfunction for $L_{\mKP},M_{\mKP}$. The compatibility condition $[L_{\mKP},M_{\mKP}]=0$ gives
\begin{align}
0 &= w_t + w_{xxx} - 6ww_xw + 3W_y + 3[w_x,W]_+ - 3[w_{xx},w] - 3[W,w^2],
\label{ncmKPequation} \\ 0 &= W_x - w_y + [w,W]. \label{ncmKPextra}
\end{align}
Equations (\ref{ncmKPequation}) and (\ref{ncmKPextra}) form a noncommutative version of the mKP equation.  Equation (\ref{ncmKPextra}) is satisfied identically by applying the change of variables \cite{MR2095463}
\begin{align}\label{variable_change}
w = - f_x f^{-1} , \quad W = - f_y f^{-1} .
\end{align}
The change of variables (\ref{variable_change}) has also been used in \cite{gilson-2008-41, Dimakis:Muller-Hoissen2006} to study a noncommutative mKP hierarchy.  For both ncKP and ncmKP equations, we consider families of solutions obtained from iterating binary Darboux transformations.  These solutions can be expressed as quasideterminants, which were introduced by Gelfand \textit{et al} in \cite{quasideterminants:main}.  An $n \times n$ matrix $Z = (z_{ij})_{n \times n}$ over a ring $\mathcal{R}$ (noncommutative, in general) has $n^2$ quasideterminants, each of which is denoted $|Z|_{ij}$ for $1 \leq i,j \leq n$.  Let $r_i^j$ denote the row vector obtained from the $i$th row of $Z$ be deleting the $j$th entry, let $ c_j^i$ denote the column vector obtained from the $j$th row of $Z$ by deleting the $i$th entry, let $Z^{ij}$ be the matrix obtained from $Z$ by deleting the $i$th row and $j$th column and assume that $Z^{ij}$ is invertible.  Then $|Z|_{ij}$ exists and
\begin{align}
|Z|_{ij} &= z_{ij} - r_i^j (Z^{ij})^{-1} c_j^i .
\end{align}
For notational convenience, we box the leading element about which the expansion is made so that
\begin{align}
|Z|_{ij} &= \begin{vmatrix} Z^{ij} & c_j^i \\ r_i^j & \fbox{$z_{ij}$} \end{vmatrix}.
\end{align}
The quasideterminant solutions obtained from binary Darboux transformations reduce to ratios of grammian determinants in the commutative limit and we call them quasigrammians.

In this paper, we will consider the case where the dependent variables in ncKP and ncmKP are matrices and apply the methods used in \cite{MR1858198, MR1634885, MR2153338, MR1799017}.  From this platform, we investigate the interaction of the two-soliton solution of the matrix versions of ncKP and ncmKP.

\section{Quasigrammian solutions of the ncKP equation}
In this section we recall the construction of the quasigrammian
solutions of ncKP in \cite{claire:jon}.  The adjoint Lax pair
is
\begin{align}
 L_{\KP}^{\dag} &= \partial_x^2 + v_x^{\dag} + \partial_y, \\
 M_{\KP}^{\dag} &= -4 \partial_x^3 - 6 v_x^{\dag} \partial_x - 3 v_{xx}^{\dag} + 3 v_y^{\dag} - \partial_t.
\end{align}
Following the standard construction of a binary Darboux
transformation (see \cite{MR1146435}), one introduces a potential $\Omega(\phi,\psi)$
satisfying
\begin{align*}
\Omega(\phi,\psi)_x &= \psi^{\dag} \phi, \quad \Omega(\phi,\psi)_y = \psi^{\dag} \phi_x - \psi_x^{\dag} \phi, \\ \Omega(\phi,\psi)_t &= -4(\psi^{\dag} \phi_{xx} - \psi_x^{\dag} \phi_x + \psi_{xx}^{\dag} \phi) - 6 \psi^{\dag} v_x \phi.
\end{align*}
The parts of this definition are compatible when $L_{\KP}[\phi] =
M_{\KP}[\phi] =0$ and $L_{\KP}^{\dag}[\psi] = M_{\KP}^{\dag}[\psi] =0$.  Note
that we can define $\Omega(\Phi, \Psi)$ for any row vectors
$\Phi$ and $\Psi$ such that $L_{\KP}[\Phi] = M_{\KP}[\Phi] =0$ and
$L_{\KP}^{\dag}[\Psi] = M_{\KP}^{\dag}[\Psi] =0$.  Consequently, if $\Phi$
is an $m$-vector and $\Psi$ is an $n$-vector, then $\Omega$ is
an $m \times n$ matrix.

A binary Darboux transformation is defined by
\begin{align*}
 \phi_{[n+1]} &= \phi_{[n]} - \theta_{[n]} \Omega(\theta_{[n]}, \rho_{[n]})^{-1} \Omega(\phi_{[n]}, \rho_{[n]})
\end{align*}
and
\begin{align*}
\psi_{[n+1]} &= \psi_{[n]} - \rho_{[n]} \Omega(\theta_{[n]}, \rho_{[n]})^{-\dag} \Omega(\theta_{[n]}, \psi_{[n]})^{\dag},
\end{align*}
in which
\[
\theta_{[n]} = \phi_{[n]}|_{\phi \rightarrow \theta_n}, \quad \rho_{[n]} = \psi_{[n]}|_{\psi \rightarrow \rho_n}.
\]
Using the notation $\Theta = (\theta_1, \ldots \theta_n)$ and $\Rho = (\rho_1, \ldots, \rho_n)$ we have, for $n \geq 1$
\begin{align*}
\phi_{[n+1]} &= \begin{vmatrix} \Omega(\Theta, \Rho) & \Omega(\phi, \Rho) \\ \Theta & \fbox{$\phi$} \end{vmatrix}, \quad \psi_{[n+1]} = \begin{vmatrix} \Omega(\Theta, \Rho)^{\dag} & \Omega(\Theta, \psi)^{\dag} \\ \Rho & \fbox{$\psi$} \end{vmatrix}
\end{align*}
and
\begin{align*}
\Omega(\phi_{[n+1]}, \psi_{[n+1]}) &= \begin{vmatrix} \Omega(\Theta, \Rho) & \Omega(\phi, \Rho) \\ \Omega(\Theta,\psi) & \fbox{$\Omega(\phi,\psi)$} \end{vmatrix}.
\end{align*}
The effect of the binary Darboux transformation
\[
\widehat{L}_{\KP} = G_{\theta, \phi}L_{\KP}G_{\theta, \phi}^{-1}, \quad  \widehat{M}_{\KP} = G_{\theta, \phi}M_{\KP}G_{\theta, \phi}^{-1}
\]
is that
\begin{align*}
 \hat{v} &= v_0 + 2 \theta \Omega(\theta, \rho)^{-1} \rho^{\dag}.
\end{align*}
After $n$ binary Darboux transformations we have
\begin{align}
 v_{[n+1]} &= v_0 + 2 \sum_{k=1}^n \theta_{[k]} \Omega(\theta_{[k]}, \rho_{[k]})^{-1} \rho_{[k]}^{\dag} \nonumber \\ &= v_0 - 2 \begin{vmatrix} \Omega(\Theta, \Rho) & \Rho^{\dag} \\ \Theta & \fbox{$0$} \end{vmatrix}.
\end{align}

\subsection{Two-soliton matrix solution}

We now derive matrix solutions of ncKP using the methods
applied in \cite{MR1858198}.  The trivial vacuum solution $v_0=\textrm{O}$ gives
\begin{align}
 v &= -2 \begin{vmatrix} \Omega(\Theta, \Rho) & \Rho^{\dag} \\ \Theta & \fbox{$0$} \end{vmatrix}. \label{vacKPsol}
\end{align}
The eigenfunctions $\theta_i$ and the adjoint eigenfunctions $\rho_i$ satisfy
\begin{align}\label{linear_dispersion_eqns}
\theta_{i,xx} = \theta_{i,y},  \quad \theta_{i,t} = -4 \theta_{i,xxx},
\end{align}
and
\begin{align}\label{binary_linear_dispersion_eqns}
\rho_{i,xx} = - \rho_{i,y},  \quad \rho_{i,t} = -4 \rho_{i,xxx},
\end{align}
respectively.  We choose the simplest nontrivial solutions of
(\ref{linear_dispersion_eqns}) and
(\ref{binary_linear_dispersion_eqns}):
\begin{align}\label{dispersion_eqn_sols}
 \theta_j &= A_je^{\eta_j}, \quad \rho_i = B_i e^{-\gamma_i},
\end{align}
where $\eta_j = p_j(x+p_jy -4p_j^2t), \gamma_i = q_i(x+q_iy-4q_i^2t)$ and $A_j, B_i$ are $d \times m$ matrices.  With this, we have
\begin{align*}
 \Omega(\theta_j, \rho_i) &= \frac{B_i^T A_j}{(p_j-q_i)}e^{(\eta_j - \gamma_i)} + \delta_{i,j}I.
\end{align*}
We take $A_j = r_j P_j$, where $r_j$ is a scalar and $P_j$ is a projection matrix. With $A_j$ chosen in this way, we must have $m=d$. We choose $B_i = I$ and the solution $u$ will be a $d \times d$ matrix.

In the case $n=1$, we obtain a one-soliton matrix solution. Expanding (\ref{vacKPsol}) gives
\begin{align*}
 v &= \frac{2rP}{e^{(\gamma - \eta)} + \frac{r}{(p-q)}}.
\end{align*}
The above calculation and others that that follow use the formula \\ $(I-aP)^{-1} = I +
aP(1-a)^{-1}$ where $a \neq 1$ is a scalar and $P$ is any
projection matrix.  We now have
\begin{align}
 u &= v_x = \frac{1}{2}(p-q)^2 P \sech^2 \left(\frac{1}{2}\left(\eta-\gamma + \xi\right)\right),
\end{align}
where $\xi = \log \left(\frac{r}{(p-q)}\right)$.

In the case $n=d=2$, we obtain a two-soliton $2 \times 2$ matrix solution.  By expanding (\ref{vacKPsol}) we get
\begin{align*}
v &= 2 \begin{pmatrix}  A_1 e^{\eta_1} & A_2 e^{\eta_2} \end{pmatrix}
    \begin{pmatrix} A_j \frac{e^{(\eta_j - \gamma_i)}}{(p_j - q_i)} + \delta_{i,j} I \end{pmatrix}_{2 \times 2}^{-1}
\begin{pmatrix} I e^{-\gamma_1} \\ I e^{-\gamma_2} \end{pmatrix} \\
    &= 2 \begin{pmatrix} K_1 e^{\gamma_1} & K_2 e^{\gamma_2} \end{pmatrix} \begin{pmatrix} I e^{-\gamma_1} \\ I e^{-\gamma_2}\end{pmatrix} = 2(K_1 + K_2).
\end{align*}
Therefore
\begin{align*}
 K_1 \left( I + \frac{r_1 e^{(\eta_1 - \gamma_1)}}{(p_1 - q_1)}P_1 \right) &= e^{(\eta_1- \gamma_1)}A_1 - \frac{e^{(\eta_1 - \gamma_1)}}{(p_1-q_2)}K_2 A_1, \\
 K_2 \left( I + \frac{r_2 e^{(\eta_2 - \gamma_2)}}{(p_2 - q_2)}P_2 \right) &= e^{(\eta_2 - \gamma_2)}A_2 - \frac{e^{(\eta_2 - \gamma_2)}}{(p_2-q_1)}K_1 A_2.
\end{align*}
We assume that the $P_j$ are the rank-$1$ projection matrices
\begin{align*}
    P_j &= \frac{\mu_j \otimes \nu_j}{(\mu_j, \nu_j)} = \frac{\mu_j \nu_j^{T}}{\mu_j^{T} \nu_j},
\end{align*}
where the $2$-vectors $\mu_j, \nu_j$ satisfy the condition $(\mu_j,\nu_j) \neq 0$.  Solving for $K_1$ and $K_2$ gives
\begin{align*}
 K_1 &= \frac{(p_2 - q_1)}{g}\left(g_2(p_1 - q_2)I - A_2\right)A_1, \\
 K_2 &= \frac{(p_1 - q_2)}{g}\left(g_1(p_2 - q_1)I - A_1\right)A_2,	
\end{align*}
where $g_i = e^{(\gamma_i -\eta_i)} + \frac{r_i}{(p_i - q_i)}$,
$g = g_1 g_2 (p_1 - q_2)(p_2 - q_1) - \alpha r_1 r_2$ and \\
$\alpha =
\frac{(\mu_1,\nu_2)(\mu_2,\nu_1)}{(\mu_1,\nu_1)(\mu_2,
\nu_2)}$.

We now investigate the behaviour of $v_{[3]}$ as $t \rightarrow
\pm \infty$. This will demonstrate that each soliton emerges
from interaction undergoing a phase shift and that the
amplitude of each soliton may also change due to the
interaction.  We first fix $\gamma_1 - \eta_1$ and assume
without loss of generality that $0 > p_2 > q_2 > p_1 > q_1$. As
$t \rightarrow - \infty$
\begin{align*}
 v & \sim 2 \frac{r_1 P_1}{g_1}
\end{align*}
and therefore
\begin{align}
 u = v_x & \sim \frac{1}{2}(p_1-q_1)^2 P_1 \sech^2 \left(\frac{1}{2}\left(\eta_1-\gamma_1 + \xi_1^- \right)\right),
\end{align}
where $\xi_1^- = \log \left( \frac{r_1}{(p_1-q_1)} \right)$.

Note that $u= v_x$ is invariant under the transformation $v \rightarrow v + C$, where $C$ is a constant matrix.  As $t \rightarrow + \infty$, we get
\begin{align*}
    v & \! \sim \! 2 \frac{(r_2 (p_1 \! - \! q_2) - (p_2 \! - \! q_2)A_2)(p_2 \! - \! q_1)A_1 \! - \! ((p_1 \! - \! q_2)A_1 \! - \! \alpha r_1 (p_2 \! - \! q_2))(p_2 \! - \! q_2)A_2}{r_2 (p_1 - q_2)(p_2 - q_1) \left(g_1 - \frac{\alpha r_1 (p_2 - q_2)}{(p_1-q_2)(p_2-q_1)}\right)} \\ & \sim 2 \frac{\hat{r_1} \hat{P_1}}{e^{\gamma_1 - \eta_1} + \frac{p_1 \hat{r}_1}{(p_1 - q_1)}},
\end{align*}
where $\hat{r}_1 = r_1 \left( 1 - \frac{\alpha (p_1 - q_1)(p_2 -q_2)}{(p_1 -q_2)(p_2 -q_1)} \right)= \frac{r_1 (\hat{\mu}_1, \hat{\nu}_1)}{(\mu_1 , \nu_1)}$, $\hat{\mu}_1 = \mu_1 - \frac{(p_2 - q_2) (\mu_1, \nu_2)\mu_2}{(p_1 - q_2)(\mu_2, \nu_2)}$, $\hat{\nu}_1 = \nu_1 - \frac{(p_2 - q_2)(\mu_2, \nu_1)\nu_2}{(p_2 - q_1)(\mu_2, \nu_2)}$ and $\widehat{P}_1 = \frac{\hat{\mu}_1 \otimes \hat{\nu}_1}{(\hat{\mu}_1, \hat{\nu}_1)}$.  Therefore
\begin{align}
    u = v_x & \sim \frac{1}{2}(p_1-q_1)^2 \widehat{P}_1 \sech^2 \left(\frac{1}{2}\left(\eta_1-\gamma_1 + \xi_1^+ \right)\right)
\end{align}
where $\xi_1^+ = \log \left( \frac{\hat{r}_1}{(p_1 - q_1)} \right)$.

Similarly, fixing $\gamma_2 - \eta_2$ gives
\begin{align}
    u & \sim \frac{1}{2}(p_2-q_2)^2 \widehat{P}_2 \sech^2 \left(\frac{1}{2}\left(\eta_2-\gamma_2 + \xi_2^- \right)\right), \quad t \rightarrow - \infty \\
    u & \sim \frac{1}{2}(p_2-q_2)^2 P_2 \sech^2 \left(\frac{1}{2}\left(\eta_2-\gamma_2 + \xi_2^+ \right)\right), \quad t \rightarrow + \infty,
\end{align}
where $\hat{r}_2 = r_2\left( 1 - \frac{\alpha (p_1 - q_1)(p_2 -q_2)}{(p_1 -q_2)(p_2 -q_1)} \right) = \frac{r_2 (\hat{\mu}_2, \hat{\nu}_2)}{(\mu_2 , \nu_2)}$, $\hat{\mu}_2 = \mu_2 - \frac{(p_1 - q_1) (\mu_2, \nu_1)\mu_1}{(p_2 - q_1)(\mu_1, \nu_1)}$, \\ $\hat{\nu}_2 = \nu_2 - \frac{(p_1 - q_1)(\mu_1, \nu_2)\nu_1}{(p_1 - q_2)(\mu_1, \nu_1)}$, $\widehat{P}_2 = \frac{\hat{\mu}_2 \otimes \hat{\nu}_2}{(\hat{\mu}_2, \hat{\nu}_2)}$, $\xi_2^- = \log \left( \frac{\hat{r}_2}{(p_2-q_2)} \right)$ and $\xi_2^+ = \log \left( \frac{r_2}{(p_2-q_2)} \right)$.

The soliton phase shifts $\Delta_j = \xi_j^+ - \xi_j^-$ are
\[
\Delta_1 = \log\left(\frac{\hat{r}_1}{r_1}\right)\ = \log \beta, \quad \Delta_2 = \log\left(\frac{r_2}{\hat{r}_2}\right)\ = - \log \beta,
\]
where $\beta = 1 - \frac{\alpha (p_1 - q_1)(p_2 -q_2)}{(p_1 -q_2)(p_2 -q_1)}$.

The matrix amplitude of the first soliton changes from $\frac{1}{2}(p_1-q_1)^2 P_1$ to $\frac{1}{2}(p_1-q_1)^2 \widehat{P}_1$ and the matrix amplitude of the second soliton changes from $\frac{1}{2}(p_2-q_2)^2 \widehat{P}_2$ to $\frac{1}{2}(p_2-q_2)^2 P_2$ as $t$ changes from $- \infty$ to $+ \infty$.  If \\ $(\mu_1, \nu_2)=0 \, (P_2P_1 = 0)$ or $(\mu_2, \nu_1) =0 \, (P_1 P_2 =0)$, then $\alpha=0$ and therefore $\beta=1$, so there is no phase shift but the matrix amplitudes may still change.  If $(\mu_1, \nu_2)=0$ and $(\mu_2, \nu_1) =0$ (giving $P_1 P_2 = P_2P_1 = 0$), there is no phase shift or change in amplitude and so the solitons have trivial interaction.  Figure 1 shows a plot of the interaction with $P_1 = \begin{pmatrix} 1 & -2 \\ 0 & 0\end{pmatrix}$ and $P_2 = \frac{1}{121} \begin{pmatrix} 96 & -16 \\ -150 & -25 \end{pmatrix}$.

\begin{figure}[h!]
\centering
\includegraphics[width=110mm, height=110mm]{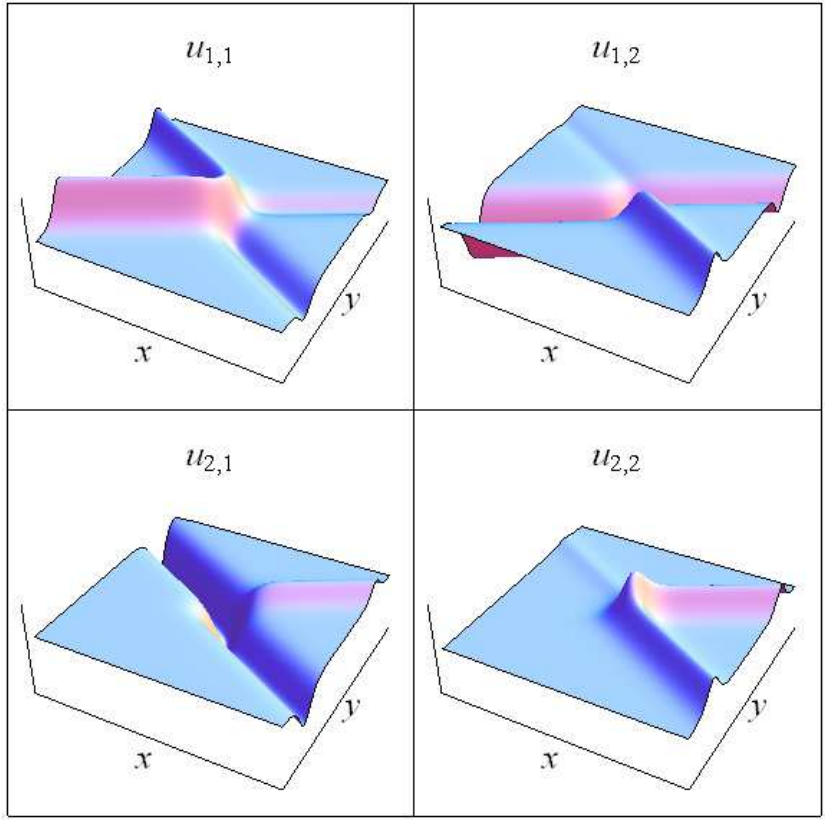}
\caption{Plot of $u=(u_{i,j})_{2 \times 2}$ with $t=0$, $p_1 = -\frac{1}{4}$, $q_1 = -\frac{39}{4}$, $p_2 = \frac{19}{2}$, $q_2= \frac{1}{2}$, $r_1 = 2$ and $r_2 =1$.}
\end{figure}

\section{Quasigrammian solutions of the ncmKP equation}

The construction of this particular binary Darboux transformation is given in \cite{Oevel:Rogers} and also in \cite{MR1786048} (for Lax operators with matrix coefficients).  The adjoint Lax pair is
\begin{align*}
    L_{\mKP}^{\dag} &= \partial_x^2 - 2w_x^{\dag} - 2 w^{\dag} \partial_x + \partial_y,
\\
    M_{\mKP}^{\dag} &= -4 \partial_x^3 + 12w^{\dag} \partial_x^2 + 6(3w^{\dag}_x \! - \!
{w^{\dag}}^2 \! \! - \! W^{\dag})\partial_x + 6(w^{\dag}_{xx} \! - \! [w^{\dag}_x,
w^{\dag}]_+ \! - \! W^{\dag}_x) -\partial_t.
\end{align*}
For notational convenience, we denote an
element of $\ker L_{\mKP}^{\dag} \cap \ker M_{\mKP}^{\dag}$ by $\phi_x$.  One introduces a potential $\Omega(\phi,\psi)$ satisfying
\begin{align*}
\Omega(\phi,\psi)_x & = \psi^{\dag} \phi_x, \quad \Omega(\phi,\psi)_y = 2 \psi^{\dag}w \phi_x + \psi^{\dag}\phi_{xx} - \psi^{\dag}_x \phi_x, \\
\Omega(\phi,\psi)_t &= 2(-2 \psi_{xx}^{\dag} \phi_x - 2 \psi^{\dag} \phi_{xxx} + 2\psi_x^{\dag} \phi_{xx} - 3 \psi^{\dag} w^2 \phi_x - 3\psi^{\dag} W \phi_x - 3\psi^{\dag} w_x \phi_x \\ & \qquad + 6 \psi_x^{\dag}w \phi_x - 6 \psi^{\dag} w \phi_{xx}).
\end{align*}
A binary Darboux transformation is defined by
\begin{align*}
\phi_{[n+1]} &= \phi_{[n]} - \theta_{[n]} \Omega(\rho_{[n]},
\theta_{[n]})^{-1} \Omega(\rho_{[n]}, \phi_{[n]})
\end{align*}
and
\begin{align*}
\psi_{[n+1]} &= \psi_{[n]} - \rho_{[n]} \Omega(\rho_{[n]},
\theta_{[n]})^{\dag^{-1}} \Omega(\psi_{[n]}, \theta_{[n]})^{\dag},
\end{align*}
in which
\[
\theta_{[n]} = \phi_{[n]}|_{\phi \rightarrow \theta_n}, \quad \rho_{[n]} = \psi_{[n]}|_{\psi \rightarrow \rho_n}.
\]
Using the notation $\Theta = (\theta_1, \ldots \theta_n)$ and $\Rho = (\rho_1, \ldots, \rho_n)$, we have, for $n \geq 1$
\begin{align*}
\phi_{[n+1]} &= \begin{vmatrix} \Omega(\Theta, \Rho) & \Omega(\phi, \Rho) \\ \Theta & \fbox{$\phi$} \end{vmatrix}, \quad \psi_{[n+1]} = \begin{vmatrix} \Omega(\Theta, \Rho)^{\dag} & \Omega(\Theta, \psi)^{\dag} \\ \Rho & \fbox{$\psi$} \end{vmatrix}.
\end{align*}
The effect of the binary Darboux transformation
\[
\widehat{L}_{\mKP} = G_{\theta, \phi_x} L_{\mKP} G_{\theta, \phi_x}^{-1}, \quad \widehat{M}_{\mKP} = G_{\theta, \phi_x} M_{\mKP} G_{\theta, \phi_x}^{-1}
\]
is that
\begin{align*}
\hat{f} &= \begin{vmatrix} \Omega & \rho^{\dag} \\ \theta & \fbox{$1$}
\end{vmatrix}f,
\end{align*}
where $f$ is given in (\ref{variable_change}).  After $n$ Darboux transformations we have
\begin{align*}
f_{[n+1]} &= \begin{vmatrix} \Omega(\Rho, \Theta) & \Rho^{\dag} \\ \Theta & \fbox{$I$} \end{vmatrix} f.
\end{align*}

\subsection{Two-soliton matrix solution}

The trivial vacuum solution $f=I$ (giving $w=W=\textrm{O}$) gives
\begin{align}
F &= \begin{vmatrix} \Omega(\Rho, \Theta) & \Rho^{\dag} \\ \Theta &
\fbox{$I$} \end{vmatrix}. \label{vacmKPsol}
\end{align}
The eigenfunctions $\theta_i$ and the adjoint eigenfunctions
$\rho_i$ satisfy (\ref{linear_dispersion_eqns}) and
(\ref{binary_linear_dispersion_eqns}) respectively.  We again choose the eigenfunction solutions of the form (\ref{dispersion_eqn_sols}). With
this, we have
\begin{align*}
 \Omega(\theta_j, \rho_i) &= \delta_{i,j}I - \frac{p_j B_i^T A_j}{q_i(p_j-q_i)}e^{(\eta_j - \gamma_i)}.
\end{align*}
As in the previous section, we take $A_j = r_j P_j$  and $B_i =
I$.  So the solutions $w$ and $W$ will be $d \times d$ matrices.

In the case $n=1$, we obtain a one-soliton matrix solution.  Expanding (\ref{vacmKPsol}) gives
\begin{align*}
 F &= I + \frac{\frac{r}{q}P}{e^{(\gamma - \eta)} - \frac{rp}{q(p-q)}}.
\end{align*}
If $r>0$ and either $q>p>0$ or $0>q>p$, or alternatively, if $r<0$ and either $p>q>0$ or $0>p>q$ then
\begin{align*}
    w &= -F_x F^{-1} = -\frac{1}{4} (pq)^{-\frac{1}{2}} (p-q)^2 P \sech\left(\frac{1}{2} \left(\eta - \gamma + \varphi \right ) \right) \sech\left( \frac{1}{2} \left ( \eta - \gamma + \chi \right ) \right),\\
    W &= -F_y F^{-1} = (p+q)w,
\end{align*}
where $\varphi = \log(\frac{-pr}{q(p-q)})$ and $\chi = \log(\frac{-r}{(p-q)})$. Both $w$ and $W$ have a unique maximum where
\begin{align*}
    \eta-\gamma &= -\log\left(\frac{-(pq^{-1})^{\frac{1}{2}}r}{(p-q)}\right) = \lambda.
\end{align*}

In the case $n=d=2$, we obtain a two-soliton $2 \times 2$ matrix solution.  Expanding (\ref{vacmKPsol}) gives
\begin{align*}
F &= I + \begin{pmatrix}  A_1 e^{\eta_1} & A_2 e^{\eta_2} \end{pmatrix}
    \begin{pmatrix} \delta_{i,j}I - \frac{p_j A_j}{q_i(p_j-q_i)}e^{(\eta_j - \gamma_i)} \end{pmatrix}_{2 \times 2}^{-1}
\begin{pmatrix} I \frac{e^{-\gamma_1}}{q_1} \\ I \frac{e^{-\gamma_2}}{q_2} \end{pmatrix} \\
    &= I + \begin{pmatrix} L_1 e^{\gamma_1} & L_2 e^{\gamma_2} \end{pmatrix} \begin{pmatrix} I \frac{e^{-\gamma_1}}{q_1} \\ I \frac{e^{-\gamma_2}}{q_2} \end{pmatrix} = I + \frac{1}{q_1}L_1 + \frac{1}{q_2}L_2.
\end{align*}
Solving for $L_1$ and $L_2$ gives
\begin{align*}
L_1 &= \frac{(p_2 -q_1)q_1}{h}((p_1 - q_2)q_2 h_2 I + p_1 A_2)A_1, \\
L_2 &= \frac{(p_1 -q_2)q_2}{h}((p_2 - q_1) q_1 h_1 I + p_2 A_1)A_2,
\end{align*}
where $h_i = e^{(\gamma_i -\eta_i)} - \frac{p_i r_i}{(p_i -
q_i)q_i}$, $h = h_1 h_2 q_1 q_2(p_1 - q_2)(p_2 - q_1) -
\alpha p_1 p_2 r_1 r_2$ and $\alpha$ is as defined in the previous section.

We now investigate the behaviour of
$F$ as $t \rightarrow \pm \infty$.  We first fix $\eta_1 -
\gamma_1$ and assume without loss of generality that
$0>p_2>q_2>p_1>q_1$.  Then, as $t \rightarrow -\infty$,
\begin{align*}
    F &\sim I + \frac{\frac{r_1}{q_1}P_1}{h_1}
\end{align*}
and therefore
\begin{align}
    w = -F_x F^{-1} &\sim  -\frac{1}{4} (p_1q_1)^{-\frac{1}{2}} (p_1-q_1)^2 P_1 \sech\left(\frac{1}{2} \left(\eta_1 - \gamma_1 + \varphi_1^- \right ) \right) \nonumber \\ & \qquad \times \sech\left( \frac{1}{2} \left ( \eta_1 - \gamma_1 + \chi_1^- \right ) \right),
\end{align}
where $\varphi_1^- = \log(\frac{-p_1r_1}{q_1(p_1-q_1)})$ and
$\chi_1^- = \log(\frac{-r_1}{(p_1-q_1)})$.

Note that $w=-F_x F^{-1}$ and $W=-F_y F^{-1}$ are invariant
under the transformation $F \rightarrow FC$ where $C$ is a non-singular
constant matrix.  As $t \rightarrow + \infty$, we get
\begin{align*}
    F & \! \sim \! I \! + \! \frac{(r_2 p_2 (p_1 \! - \! q_2) \!- \! p_1 (p_2 \! - \! q_2)A_2)(p_2 \! - \! q_1)A_1 \! - (p_2(p_1 \! - \! q_2)A_1 \! - \! \alpha p_1 r_1 (p_2 \! - \! q_2))}{h_1 r_2 p_2 q_1 (p_1 \! - \! q_2)(p_2 \! - \! q_1) + \alpha p_1 p_2 r_1 r_2 (p_2 \! - \! q_2)} \\ & \quad \times  (p_2 \! - \! q_2)A_2 \left(I + \frac{(p_2-q_2)A_2}{q_2r_2}\right) \\
     &\sim I + \frac{\frac{\tilde{r}_1}{q_1}\widetilde{P}_1}{e^{\gamma_1 - \eta_1} - \frac{p_1 \tilde{r}_1}{q_1 (p_1-q_1)}},
\end{align*}
where $\tilde{r}_1 = r_1 \left( 1 - \frac{\alpha (p_1 -
q_1)(p_2 -q_2)}{(p_1 -q_2)(p_2 -q_1)} \right)= \frac{r_1
(\tilde{\mu}_1, \tilde{\nu}_1)}{(\mu_1 , \nu_1)}$,
$\tilde{\mu}_1 = \mu_1 - \frac{p_1(p_2 - q_2) (\mu_1,
\nu_2)\mu_2}{p_2(p_1 - q_2)(\mu_2, \nu_2)}$, $\tilde{\nu}_1 =
\nu_1 - \frac{q_1(p_2 - q_2)(\mu_2, \nu_1)\nu_2}{q_2(p_2 -
q_1)(\mu_2, \nu_2)}$ and $\widetilde{P}_1 = \frac{\tilde{\mu}_1
\otimes \tilde{\nu}_1}{(\tilde{\mu}_1, \tilde{\nu}_1)}$.
Therefore
\begin{align}
    w = - F_x F^{-1} &\sim  -\frac{1}{4} (p_1q_1)^{-\frac{1}{2}} (p_1-q_1)^2 \widetilde{P}_1 \sech\left(\frac{1}{2} \left(\eta_1 - \gamma_1 + \varphi_1^+ \right ) \right) \nonumber \\ & \qquad \times \sech\left( \frac{1}{2} \left ( \eta_1 - \gamma_1 + \chi_1^+ \right ) \right),
\end{align}
where $\varphi_1^+ =
\log(\frac{-p_1\tilde{r}_1}{q_1(p_1-q_1)})$ and $\chi_1^+ =
\log(\frac{-\tilde{r}_1}{(p_1-q_1)})$.

Similarly, fixing $\gamma_2 - \eta_2$ gives
\begin{align}
    w &\sim  -\frac{1}{4} (p_2q_2)^{-\frac{1}{2}} (p_2-q_2)^2 \widetilde{P}_2 \sech\left(\frac{1}{2} \left(\eta_2 - \gamma_2 + \varphi_2^- \right ) \right) \nonumber \\ & \qquad \times \sech\left( \frac{1}{2} \left ( \eta_2 - \gamma_2 + \chi_2^- \right ) \right), \quad t \rightarrow - \infty,\\
    w &\sim  -\frac{1}{4} (p_2q_2)^{-\frac{1}{2}} (p_2-q_2)^2 P_2 \sech\left(\frac{1}{2} \left(\eta_2 - \gamma_2 + \varphi_2^+ \right ) \right) \nonumber \\ & \qquad \times \sech\left( \frac{1}{2} \left ( \eta_2 - \gamma_2 + \chi_2^+ \right ) \right), \quad t \rightarrow + \infty,
\end{align}
where $\tilde{r}_2 = r_2 \left( 1 - \frac{\alpha (p_1 -
q_1)(p_2 -q_2)}{(p_1 -q_2)(p_2 -q_1)} \right)= \frac{r_2
(\tilde{\mu}_2, \tilde{\nu}_2)}{(\mu_2 , \nu_2)}$,
$\tilde{\mu}_2 = \mu_2 - \frac{p_2(p_1 - q_1) (\mu_2,
\nu_1)\mu_1}{p_1(p_2 - q_1)(\mu_1, \nu_1)}$, $\tilde{\nu}_2 =
\nu_2 - \frac{q_2(p_1 - q_1)(\mu_1, \nu_2)\nu_1}{q_1(p_1 -
q_2)(\mu_1, \nu_1)}$, $\widetilde{P}_2 = \frac{\tilde{\mu}_2
\otimes \tilde{\nu}_2}{(\tilde{\mu}_2, \tilde{\nu}_2)}$,
$\varphi_2^- = \log(\frac{-p_2\tilde{r}_2}{q_2(p_2-q_2)})$,
$\chi_2^- = \log(\frac{-\tilde{r}_2}{(p_2-q_2)})$, $\varphi_2^+ = \log(\frac{-p_2 r_2}{q_2(p_2-q_2)})$ and $\varphi_2^- = \log(\frac{-p_2 r_2}{q_2(p_2-q_2)})$.

The soliton phase shifts $\Lambda_i = \lambda_i^+ - \lambda_i^-$ are
\begin{align*}
\Lambda_1 &= \log\left(\frac{r_1}{\tilde{r}_1} \right)\ = - \log \beta, \quad \Lambda_2 = \log\left(\frac{\tilde{r}_2}{r_2}\right)\ = \log \beta.
\end{align*}

The matrix amplitude of the first soliton changes from \\
$-\frac{1}{4} (p_1q_1)^{-\frac{1}{2}} (p_1-q_1)^2 P_1$ to
$-\frac{1}{4} (p_1q_1)^{-\frac{1}{2}} (p_1-q_1)^2
\widetilde{P}_1$ and the matrix amplitude of the second soliton
changes from $-\frac{1}{4} (p_2q_2)^{-\frac{1}{2}} (p_2-q_2)^2
\widetilde{P}_2$ to \\ $-\frac{1}{4} (p_2q_2)^{-\frac{1}{2}}
(p_2-q_2)^2 P_2$ as $t$ changes from $- \infty$ to $+ \infty$.
Figure 2 shows a plot of the interaction with $P_1 = \begin{pmatrix} 1 & -1 \\ 0 & 0\end{pmatrix}$ and $P_2 = \frac{1}{13} \begin{pmatrix} 16 & -6 \\ 8 & -3 \end{pmatrix}$.

\begin{figure}[h!]
\centering
\includegraphics[width=110mm, height=110mm]{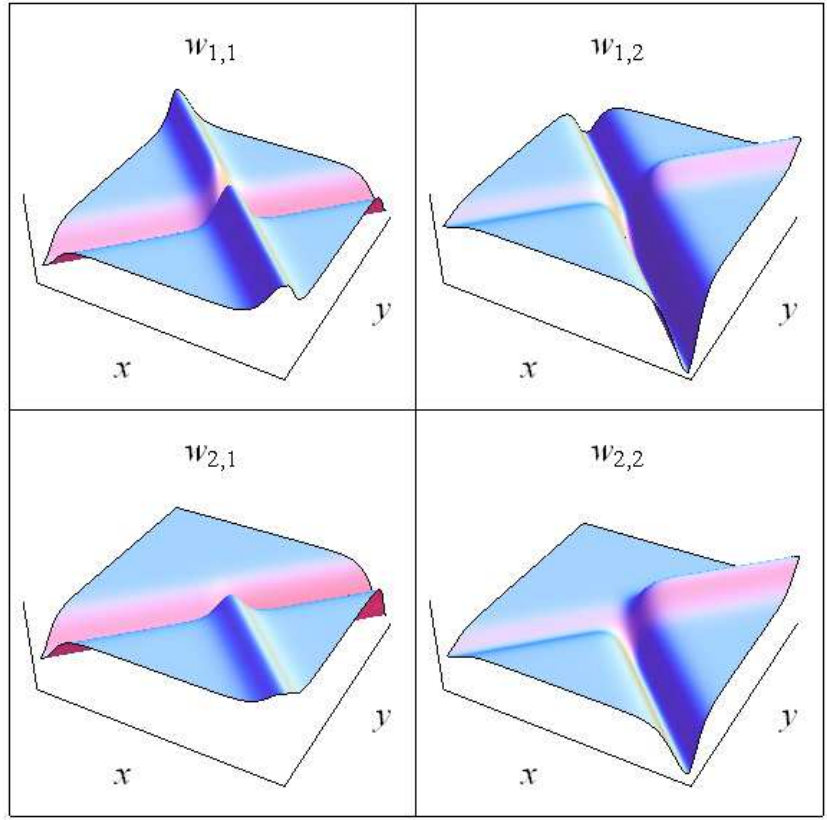}
\caption{Plot of $w=(w_{i,j})_{2 \times 2}$ with $t=0$, $p_1 = \frac{1}{4}$, $q_1 = \frac{3}{4}$, $p_2 = -\frac{1}{4}$, $q_2= -\frac{3}{4}$, $r_1 = 1$ and $r_2 =-1$.}
\end{figure}

\section{Conclusions}
In this paper, we have considered a noncommutative KP and a noncommutative mKP equation.  It was shown that solutions of ncmKP obtained from a binary Darboux transformation could be expressed as a single quasideterminant.  In addition, we have used methods similar to those employed in \cite{MR1858198} and obtained matrix versions of both ncKP and ncmKP.  Finally, we investigated the interaction properties of the two-soliton solution of both ncKP and ncmKP.  This showed that as well as undergoing a phase-shift, the amplitude of each soliton can also change, giving a more elegant picture than the commutative case.

\bibliographystyle{plain}
\bibliography{gallipoli}

\end{document}